\documentclass{emulateapj}
\usepackage{lscape}

\newcommand{\lbol}{\mbox{$L_{bol}$}} 
\newcommand{\tbol}{\mbox{$T_{bol}$}} 

\newcommand{\kospal}{K{\'o}sp{\'a}l}

\newcommand{\degree}{\mbox{$^{\circ}$}}

\newcommand{\as}{\mbox{\arcsec}}
\newcommand{\kms}{\mbox{km s$^{-1}$}}
\newcommand{\mjybeam}{\mbox{mJy beam$^{-1}$}}

\newcommand{\um}{$\mu$m}
\newcommand{\lsun}{\mbox{L$_\odot$}}
\newcommand{\msun}{\mbox{M$_\odot$}}

\newcommand{\co}{$^{12}$CO}
\newcommand{\coo}{$^{13}$CO}
\newcommand{\cooo}{C$^{18}$O}

\newcommand{\ntdp}{N$_2$D$^+$}

\newcommand{\cojtwo}{$^{12}$CO J $=2-1$}

\newcommand{\cojfourteen}{$^{12}$CO J $=14-13$}
\newcommand{\coojtwo}{$^{13}$CO J $=2-1$}

\newcommand{\cooojtwo}{C$^{18}$O J $=2-1$}

\newcommand{\ntdpjthree}{N$_2$D$^+$ J $=3-2$}

\slugcomment{\footnotesize For submission to ApJ, \today}

\begin{document}
\title {\bf Revealing The Millimeter Environment of the New FU Orionis Candidate HBC722 with the Submillimeter Array}
\author{
Michael M.~Dunham\altaffilmark{1, 2}, 
H\'ector G.~Arce\altaffilmark{1}, 
Tyler L.~Bourke\altaffilmark{3}, 
Xuepeng Chen\altaffilmark{1}, 
Tim A.~van Kempen\altaffilmark{4, 5},
\& Joel D.~Green\altaffilmark{6}
}

\altaffiltext{1}{Department of Astronomy, Yale University, P.O. Box 208101, New Haven, CT 06520, USA}

\altaffiltext{2}{michael.dunham@yale.edu}

\altaffiltext{3}{Harvard-Smithsonian Center for Astrophysics, 60 Garden Street, Cambridge, MA 02138, USA}

\altaffiltext{4}{Joint ALMA Offices, Av. Alsonso de Cordova, Santiago, Chile}

\altaffiltext{5}{Leiden Observatory, Leiden University, P.O. Box 9513, 2300 RA Leiden, The Netherlands}

\altaffiltext{6}{Department of Astronomy, The University of Texas at Austin, 1 University Station, C1400, Austin, Texas 78712--0259, USA}

\begin{abstract}
We present 230 GHz Submillimeter Array continuum and molecular line 
observations of the newly discovered FUor candidate HBC722.  We report the 
detection of seven 1.3 mm continuum sources in the vicinity of HBC722, none of 
which correspond to HBC722 itself.  
We compile infrared and submillimeter continuum 
photometry of each source from previous studies and conclude that three are 
Class 0 embedded protostars, one is a Class I embedded protostar, one is a 
Class I/II transition object, and two are either starless cores or very 
young, very low luminosity protostars or first hydrostatic cores.  
We detect a northwest-southeast outflow, 
consistent with the previous detection of such an outflow in low-resolution, 
single-dish observations, and note that its axis may be precessing.  
We show that this outflow is centered on and driven 
by one of the nearby Class 0 sources rather than HBC722, and find no conclusive 
evidence that HBC722 itself is driving an outflow.  
The non-detection of HBC722 in the 1.3 mm continuum observations suggests 
an upper limit of 0.02 \msun\ for the mass of the circumstellar disk.  This 
limit is consistent with typical T Tauri disks and with a disk that provides 
sufficient mass to power the burst.
\end{abstract}



\section{Introduction}\label{sec_intro}

FU Orionis objects (hereafter FUors) are a group of young, pre-main sequence 
stars observed to flare in brightness by $4-6$ magnitudes in the optical and 
remain bright for decades (Herbig 1977).  They are named after the prototype 
FU Orionis, which flared by about 6 magnitudes in 1936 and has remained 
in an elevated state to the present day (Wachmann 1954; Herbig 1966).  Only 
10 confirmed FUors are known to exist from direct observations of flares, 
with about another 10 identified based on similar spectral characteristics 
to the confirmed FUors (see Reipurth \& Aspin 2010 for a recent review).  
The large amplitude flares are attributed to enhanced accretion from the 
surrounding circumstellar disk (Hartmann \& Kenyon 1985), with the accretion 
rate from the disk onto the star increasing to up to $\sim 10^{-4}$ \msun\ 
yr$^{-1}$ (Hartmann \& Kenyon 1996).  
Various triggering mechanisms for the accretion bursts have 
been proposed, including interactions with binary companions (Bonnell \& 
Bastien 1992), thermal instabilities (Hartmann \& Kenyon 1996), and 
gravitational and magnetorotational instabilities (Zhu et al.~2007, 2009a, 
2009b; Vorobyov \& Basu 2010).  
FUors are especially interesting and relevant to general star 
formation studies because they may represent the late, optically visible end 
stages of episodic accretion bursts and luminosity flares through the duration 
of the embedded phase (e.g., Kenyon et al.~1990; Enoch et al.~2009; Evans et 
al.~2009; Vorobyov 2009; Dunham et al.~2010; Dunham \& Vorobyov 2012).  With 
so few bona-fide FUors, detailed observations of each one are necessary in 
order to characterize their properties and understand their place in the 
general star formation process.

In this paper, we present 230 GHz Submillimeter Array (SMA; Ho et 
al.~2004) observations of the newly discovered FUor candidate HBC722.  
As described in more detail in \S \ref{sec_hbc722} below, HBC722 is located 
within a small group of $\sim$ 10 young stars, greatly complicating the 
analysis of existing, low-resolution single-dish ground and space-based 
(sub)millimeter data.  The SMA 230 GHz continuum and molecular line 
observations presented in this paper are motivated by a need to disentagle 
the millimeter emission from the various sources in the vicinity of HBC722 in 
order to better determine its evolutionary status and physical properties.
The organization of this paper is as follows:  a brief 
summary of HBC722 is given in \S \ref{sec_hbc722}, a description of the 
observations and data reduction is provided in \S \ref{sec_obs}, the basic 
results are presented in \S \ref{sec_results}, including the continuum data 
in \S \ref{sec_results_continuum} and the CO line data in \S 
\ref{sec_results_co}, a discussion of the detected continuum sources 
is presented in \S \ref{sec_discussion_sources}, a discussion of the 
evolutionary status of HBC722 is given in \ref{sec_discussion_hbc722}, and 
a summary of our results is presented in \S \ref{sec_summary}.

\section{HBC722}\label{sec_hbc722}

HBC722, also known as V2493 Cyg, LkH$\alpha$ 188-G4, and PTF10qpf, is located 
at R.A.~= 20:58:17.03, decl.~= $+$43:53:43.4 (J2000) in the ``Gulf of Mexico'' 
region of the North American/Pelican Nebula Complex at a distance of 520 pc 
(Straizys et al.~1989; Laugalys et al.~2006).  
Prior to 2010 it was regarded as an emission-line star with 
a spectral type of K7$-$M0 and $A_{\rm V}$ of 3.4 magnitudes (Cohen \& 
Kuhi 1979).  Semkov et al.~(2010) and Miller et al.~(2011) independently 
reported a $4-5$ magnitude optical flare in HBC722.  This flare began 
sometime before 2010 May, reached peak brightness in late 2010 September, 
followed a similar rise in brightness as other FUors, and exhibited spectral 
characteristics indicative of FUors (Semkov \& Peneva 2010a; Semkov \& 
Peneva 2010b; Semkov et al.~2010; Munari et al.~2010; Leoni et al.~2010; 
Miller et al.~2011).  Prior to the flare, HBC722 featured a spectral energy 
distribution (SED) consistent with a Class II T Tauri star, with 
\lbol\ $= 0.85$ \lsun\ and infrared spectral slope $\alpha = -0.77$ 
(\kospal\ et al.~2011; Miller et al.~2011).  During the flare \lbol\ increased 
to $8.7-12$ \lsun.  After the peak brightness was reached in late 2010 
September, HBC722 entered a phase of rapid decline, decreasing by 
$\sim 0.5$ magnitudes in the optical by 2010 December (\kospal\ et al.~2011).  
Based on a linear extrapolation of this initial decline, \kospal\ et 
al.~(2011) predicted a return to quiescence by late 2011 or early 2012, much 
too quickly to be a FUor.  They also noted that the outburst \lbol\ 
and mass accretion rate implied by this \lbol\ are on the low end for FUors.  
However, unpublished photometry from the American Association of Variable 
Star Observers (AAVSO)\footnote{Available at http://www.aavso.org/} shows 
that the optical brightness of HBC722 stopped its decline in early 2011, 
remained constant throughout most of 2011 at $3-4$ magnitudes above the 
quiescent brightness, and increased by about $0.5-1$ magnitudes between 
2011 September and 2012 June.  The cessation in early 2011 of the initial 
rapid decline has been very recently confirmed by Lorenzetti et al.~(2012) 
and Semkov et al.~(2012).

HBC722 was imaged by the \emph{Spitzer Space Telescope} (Werner et al.~2004) 
at $3.6-8$ \um\ with the Infrared Array Camera (IRAC; Fazio et al.~2004) and 
$24-160$ \um\ with the  Multiband Imaging Photometer for Spitzer 
(MIPS; Rieke et al.~2004) 
as part of a large survey of the North American and Pelican Nebula Complex 
(Guieu et al.~2009; Rebull et al.~2011).  HBC722 is located within a small 
group of $\sim$ 10 young stars called the LkH$\alpha$ group by Cohen \& Kuhi 
(1979), all located within approximately $20-30$\as\ ($10400-15600$ AU at the 
assumed distance of 520 pc) of HBC722.  Several are detected in the 
mid-infrared with \emph{Spitzer} and are thus still associated with dust in 
surrounding disks and envelopes, emphasizing the complex nature of this 
region (see Figure 1 of Green et al.~[2011] and \S 
\ref{sec_discussion_sources} below).  

Green et al.~(2011) presented (sub)millimeter data on HBC722 and its 
surrounding 
region observed with the \emph{Herschel Space Observatory} (Pilbratt et 
al.~2011) and Caltech Submillimeter Observatory (CSO), including 
\emph{Herschel} $70-500$ \um\ images with $5-36$\as\ resolution, 
\emph{Herschel} $50-600$ \um\ full spectral scans, a 350 \um\ image with 9\as\ 
resolution obtained with the Submillimeter High Angular Resolution Camera II 
(SHARC-II) at the CSO, and a \cojtwo\ map with 30\as\ resolution obtained at 
the CSO.  HBC722 is detected at 70 \um\ with \emph{Herschel}; at all longer 
wavelengths the resolution is insufficient to resolve HBC722 from other, 
nearby sources.  Green et al.~noted that the $100-500$ \um\ 
\emph{Herschel} emission 
does not peak on HBC722, suggesting it is not the dominant source of 
submillimeter emission.  No emission is detected at the position of HBC722 
in higher-resolution 350 \um\ SHARC-II data, suggesting that this source is 
no longer associated with a circumstellar envelope, consistent with the 
pre-outburst classification as a Class II T Tauri star.  The 30\as\ resolution 
\cojtwo\ map presented by Green et al.~clearly shows a NW-SE outflow centered 
near HBC722 (see their Figure 2 and \S \ref{sec_results_co} below), but the 
resolution is insufficient to definitively identify the driving source.  In 
this paper, we present high angular resolution 1.3 mm continuum and \cojtwo\ 
observations obtained with the SMA in order to identify which sources in 
the vicinity of HBC722 are associated with (sub)millimeter continuum 
emission, characterize each source, and identify the driving source(s) of the 
outflow(s) in this region.

\section{Observations and Data Reduction}\label{sec_obs}

One track of observations of HBC722 was obtained with the SMA on 20 May 
2011 in the compact configuration with seven antennas, providing projected 
baselines ranging from 5 to 76 m.  A two pointing mosaic was adopted to 
fully map the multiple sources in the vicinity of HBC722 (see, e.g., Figure 1 
of Green et al.~[2011]).  The phase centers of the two pointings are 
R.A.=20:58:16.56, decl.=$+$43:53:52.9 (J2000) and R.A.=20:58:17.67, 
decl.=$+$43:53:31.0 (J2000), with the total observing time divided equally 
between the two pointings.  The observations were obtained with the 230 GHz 
receiver and included 4 GHz of bandwidth per sideband, with 12 GHz spacing 
between the centers of the two sidebands.  The correlator was configured such 
that the lower sideband (LSB) covered approximately $216.8-220.8$ GHz while 
the upper sideband (USB) covered aproximately $228.8-232.8$ GHz, providing 
simultaneous observations of the \co, \coo, \cooojtwo, \ntdpjthree, and SiO 
J $=5-4$ lines.  The lines were observed with either 256 (\coo, SiO) or 512 
(\co, \cooo, \ntdp) channels in the 104 MHz bands, providing channel 
separations of 0.53 and 0.26 \kms, respectively.  The remaining bands were 
used to measure the 1.3 mm continuum with a total bandwidth of 5.49 GHz.

The observations were obtained in moderate weather conditions, with the 
zenith opacity at 225 GHz ranging between $0.2-0.25$ and the system 
temperature typically $\sim$ 250 K, ranging between $150-400$ K depending on 
elevation.  Regular observations of the sources MWC349a and BLLAC were 
interspersed with those of HBC722 for gain calibration.  Saturn and 3C279 were 
used for passband calibration, and Uranus was used for absolute flux 
calibration.  We estimate a 20\% uncertainty in the absolute flux calibration 
by comparing the measured fluxes of the calibrators from our calibrated data 
with those in the SMA calibrator database\footnote{Available at 
http://sma1.sma.hawaii.edu/callist/callist.html} for the same observation date.

\begin{deluxetable*}{lcccccc}
\tabletypesize{\scriptsize}
\tablewidth{0pt}
\tablecaption{\label{tab_smaobs}SMA Observations of HBC722}  
\tablehead{
\colhead{} & \colhead{$\nu$} & \colhead{Beam FWHM} & \colhead{Beam PA\tablenotemark{a}} & \colhead{Bandwidth} & \colhead{$\delta$V\tablenotemark{b}} & \colhead{$1\sigma$ rms\tablenotemark{c}}\\
\colhead{Line} & \colhead{(GHz)} & \colhead{(Arcseconds)} & \colhead{(Degrees)} & \colhead{(GHz)} & \colhead{(\kms)} & \colhead{(\mjybeam)}}
\startdata
\cojtwo\ & 230.53797 & $2.93 \times 2.74$ & $-$50.5 & 0.082 & 0.5 & 126 \\
\coojtwo\ & 220.39868 & $3.15 \times 2.81$ & $-$46.0 & 0.082 & 0.5 & 103 \\
\hline
Continuum & 225.44882 & $2.91 \times 2.74$ & $-$52.1 & 5.494 & \nodata & 1.65 
\enddata
\tablenotetext{a}{Position angle of the long axis of the beam, measured east (counterclockwise) from north.}
\tablenotetext{b}{Width of each channel in \kms.}
\tablenotetext{c}{For lines, the mean of the 1$\sigma$ rms of the spectrum at each spatial position with the spectral resolution given in the previous column.  For the continuum, the $1\sigma$ rms of the continuum intensity.}
\end{deluxetable*}

\begin{figure}
\epsscale{1.0}
\plotone{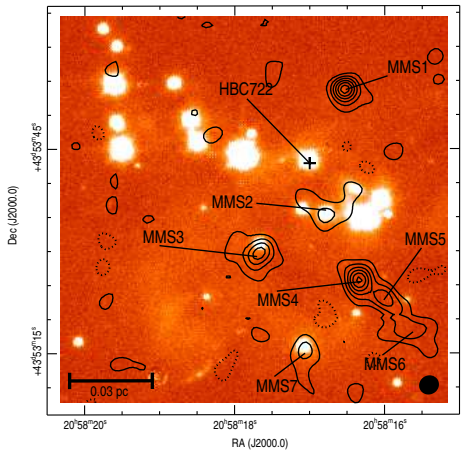}
\caption{\label{fig_cont}UKIDSS K$_{\rm s}$ band image of the vicinity of 
HBC722, taken from Green et al.~(2011).  The cross marks the position of 
HBC722 itself.  Overplotted with solid black lines are SMA 1.3 mm emission 
contours starting at 2$\sigma$ and increasing by 2$\sigma$, where the 
1$\sigma$ rms in the continuum image is 1.65 mJy beam$^{-1}$.  Also plotted 
with dotted black lines are SMA 1.3 mm emission contours starting at 
$-$2$\sigma$ and decreasing by 2$\sigma$.  The 
synthesized beam size and shape of the SMA 1.3 mm continuum observations is 
shown by the black filled oval in the lower right.}
\end{figure}

The data were inspected, flagged, and calibrated using the MIR software 
package\footnote{Available at 
https://www.cfa.harvard.edu/$\sim$cqi/mircook.html} and imaged, cleaned, and 
restored using the Multichannel Image Reconstruction, Image Analysis, and 
Display (MIRIAD) software package configured for the SMA\footnote{Available at 
http://www.cfa.harvard.edu/sma/miriad/}.  Since our data are a mosaic of two 
pointings, we originally tried to use the MIRIAD clean task 
\emph{mossdi} appropriate for mosaics.  However, \emph{mossdi} only allows for 
a single cleaning iteration, and since the region surrounding HBC722 
shows complicated morphology, especially in the \co\ data (see \S 
\ref{sec_results} below), we found the result to be unacceptable since 
emission from the sidelobes of the dirty beam were very obviously present in 
our final image independent of the exact cleaning parameters adopted.  Thus, 
we instead cleaned and imaged each of the two pointings separately using 
the MIRIAD task \emph{clean}, using an iterative cleaning process where we 
first cleaned only those regions showing clear emission in the dirty maps and 
then used those results as input models to further passes of \emph{clean} 
applied over the full images.   In the first iteration the regions varied from 
one channel to the next in the molecular line data since the emission 
morphology is different at different velocities.  Finally, we mosaiced 
together the final images from each of the two pointings.  
Both pointings were corrected for primary beam attenuation before 
mosaicing.  While all quantitative results in this paper are derived from the 
mosaics created after such correction, for display purposes the images and 
figures are created from the mosaics created without such corrections, unless 
otherwise indicated.  Imaging was performed with a robust uv weighting 
parameter of $+$1.  The final maps were re-gridded onto 0.5\as\ pixels.

Only the continuum and \co\ and \coo\ are detected and discussed in this 
paper.  Table \ref{tab_smaobs} lists, for each of these observations, 
the frequency of observation, the synthesized beam size and orientation, the 
total bandwidth, the channel separation (for the lines), and the measured 
$1\sigma$ rms.  For the continuum, the 1$\sigma$ rms is determined by 
calculating the standard deviation of all off-source pixels.  For the spectra, 
the 1$\sigma$ rms is determined by calculating, for each pixel, the standard 
deviation of the intensity in each spectral channel outside of the velocity 
range of $-$20 to 30 \kms, and then calculating the mean standard deviation 
over all pixels.

\section{Results}\label{sec_results}

\subsection{Continuum}\label{sec_results_continuum}

Figure \ref{fig_cont} displays the UKIDSS K$_{\rm s}$ band image of the 
vicinity of HBC722, taken from Green et al.~(2011).  Overlaid are contours 
showing the SMA 1.3 mm continuum intensity.  
Seven individual millimeter sources are detected and 
labeled as MMS1 -- MMS7 in order of decreasing declination.  As evident from 
this Figure, HBC722 itself is not detected at 1.3 mm.  
Associations between the seven millimeter sources detected by 
the SMA and known sources at infrared and submillimeter wavelengths will be 
discussed in \S \ref{sec_discussion_sources}, and the non-detection of 
HBC722 itself will be discussed in \S \ref{sec_discussion_hbc722}.

\begin{deluxetable*}{lcccccccc}
\tabletypesize{\scriptsize}
\tablewidth{0pt}
\tablecaption{\label{tab_continuum}Elliptical Gaussian Fits to the 1.3 mm Continuum Detections}  
\tablehead{
\colhead{} & \colhead{Peak R.A.} & \colhead{Peak Decl.} & \colhead{Peak Flux Density\tablenotemark{a}} & \colhead{Total Flux Density\tablenotemark{b}} & \colhead{Source Size\tablenotemark{c}} & \colhead{Source P.A.\tablenotemark{c}} & \colhead{$M$\tablenotemark{d}} & \colhead{$n$\tablenotemark{d}} \\
\colhead {Source} & \colhead{(J2000)}   & \colhead{(J2000)}    & \colhead{(mJy beam$^{-1}$)} & \colhead{(mJy)}      & \colhead{(\as)}       & \colhead{(degrees)}   & \colhead{(\msun)} & \colhead{(cm$^{-3}$)}}
\startdata
MMS1                  & 20:58:16.52 & $+$43:53:53.7 & 27.4 $\pm$ 6.0 & 38.3 $\pm$  7.7 & 2.0 $\times$ 1.6 & $+$26.6 & 0.15 & 5.3 $\times$ 10$^7$ \\
MMS2                  & 20:58:16.78 & $+$43:53:36.0 &  7.7 $\pm$ 2.1 & 55.1 $\pm$ 11.0 & 8.3 $\times$ 6.1 & $-$70.4 & 0.21 & 1.2 $\times$ 10$^6$ \\
MMS3                  & 20:58:17.70 & $+$43:53:29.2 & 18.6 $\pm$ 4.1 & 66.1 $\pm$ 13.2 & 4.9 $\times$ 4.3 & $+$31.8 & 0.25 & 5.3 $\times$ 10$^6$ \\
MMS4                  & 20:58:16.32 & $+$43:53:25.5 & 33.2 $\pm$ 7.1 & 60.5 $\pm$ 12.1 & 3.3 $\times$ 1.8 & $+$30.2 & 0.23 & 3.2 $\times$ 10$^7$ \\
MMS5                  & 20:58:16.01 & $+$43:53:22.8 & 25.1 $\pm$ 5.5 & 62.7 $\pm$ 12.5 & 5.9 $\times$ 1.0 & $+$55.6 & 0.24 & 3.4 $\times$ 10$^7$ \\
MMS6                  & 20:58:15.62 & $+$43:53:18.1 & 15.5 $\pm$ 3.8 & 44.3 $\pm$  8.9 & 6.5 $\times$ 1.6 & $+$89.4 & 0.17 & 1.0 $\times$ 10$^7$ \\
MMS7                  & 20:58:17.06 & $+$43:53:15.0 & 11.0 $\pm$ 2.5 & 20.3 $\pm$  4.1 & 3.1 $\times$ 2.1 &  $+$1.2 & 0.08 & 1.0 $\times$ 10$^7$
\enddata
\tablenotetext{a}{Uncertainties include the statistical uncertainty returned by \emph{imfit} and the 20\% calibration uncertainy added in quadrature.}
\tablenotetext{b}{Uncertainties include only the 20\% calibration uncertainty since \emph{imfit} does not return a statistical uncertainty for this parameter.}
\tablenotetext{c}{Deconvolved with the beam (see Table \ref{tab_smaobs} for the beam size and shape).  The source sizes for the weakest two sources (MMS2 and MMS7) are likely quite uncertain and are best treated as estimates only.}
\tablenotetext{d}{No uncertainties are given for the mass and mean number density since they are dominated by the uncertain assumptions for the dust temperature and opacity.  See the text in \S \ref{sec_results_continuum} for more details.}
\end{deluxetable*}

\begin{figure*}
\epsscale{1.2}
\plotone{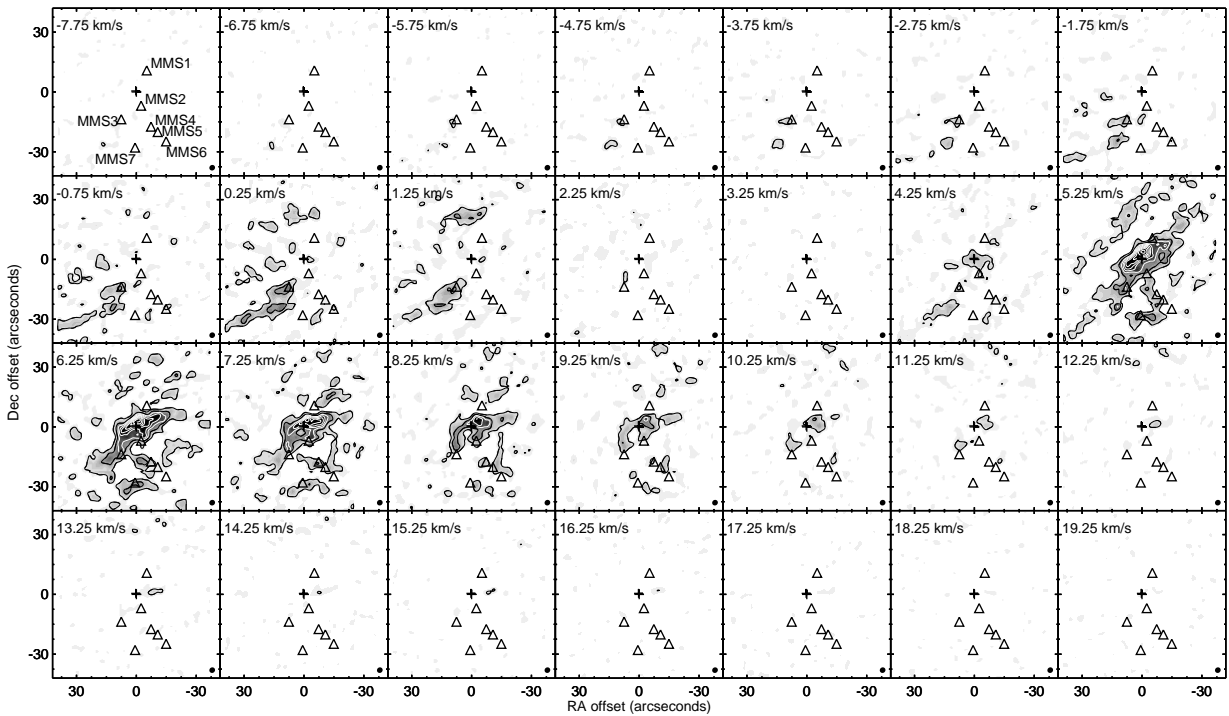}
\caption{\label{fig_12co21_cmap}SMA \cojtwo\ channel maps of the region 
surrounding HBC722 with a velocity resolution of 1 \kms, with the velocities 
marked in the top left of each panel.  The central (0,0) position is that of 
HBC722 itself and is marked with a cross in each panel.  
The positions of the 7 sources detected in the SMA 1.3 mm continuum data 
are plotted as triangles and labeled in the first panel.  
The gray scale shows intensity in units of Jy beam$^{-1}$ with 
a linear scaling ranging from -0.05 Jy beam$^{-1}$ (white) to 4.0 Jy 
beam$^{-1}$ (black).  The solid contours start at 5$\sigma$ and increase by 
10$\sigma$, where the 1$\sigma$ rms is 0.094 Jy beam$^{-1}$ in the 1 \kms\ 
channels.  The lowest two contours are plotted in black; all others are 
plotted in white.  The dotted contours start at $-$5$\sigma$ and decrease by 
10$\sigma$.  The synthesized beam size and shape of the \cojtwo\ observations 
is shown by the black filled oval in the lower right of each panel.}
\end{figure*}

For all seven sources, we used the MIRIAD task \emph{imfit} to fit 
an elliptical Guassian to each source.  The properties of each source as 
derived from these fits, including the peak position, peak flux density, 
total flux density, deconvolved source size, and deconvolved position angle, 
are reported in Table \ref{tab_continuum}.  The last two columns of Table 
\ref{tab_continuum} list the mass and density of each source derived from the 
SMA continuum detections.  The mass is calculated as

\begin{equation}\label{eq_dustmass}
M = 100 \frac{d^2 S_{\nu}}{B_{\nu}(T_D) \kappa_{\nu}} \quad ,
\end{equation}
where $S_{\nu}$ is the total flux density, $B_{\nu}(T_D)$ is the Planck 
function at the isothermal dust temperature $T_D$, $\kappa_{\nu}$ is the dust 
opacity, $d = 520$ pc, and the factor of 100 is the assumed gas-to-dust ratio.  
We adopt the dust opacities of Ossenkopf \& Henning (1994) appropriate for 
thin ice mantles after $10^5$ yr of coagulation at a gas density of $10^6$ 
cm$^{-3}$ (OH5 dust), giving $\kappa_{\nu} = 0.867$ cm$^2$ gm$^{-1}$ at the 
frequency of the continuum observations.  To calculate the mass we assume 
$T_D = 30$ K.  No uncertainties are given for the masses because these 
uncertainties are dominated by uncertain assumptions for the dust temperature 
and opacity.  For example, changing the assumed $T_D$ to 10 K would increase 
the masses by a factor of 4.5.  Furthermore, the dust opacity at 1.3 mm can 
vary by factors of $\sim 2-4$ depending on which dust opacity model is adopted 
(e.g., Shirley et al.~2005; 2011), directly leading to factors of $\sim 2-4$ 
variation in the mass.

\begin{figure}[hbt!]
\epsscale{1.0}
\plotone{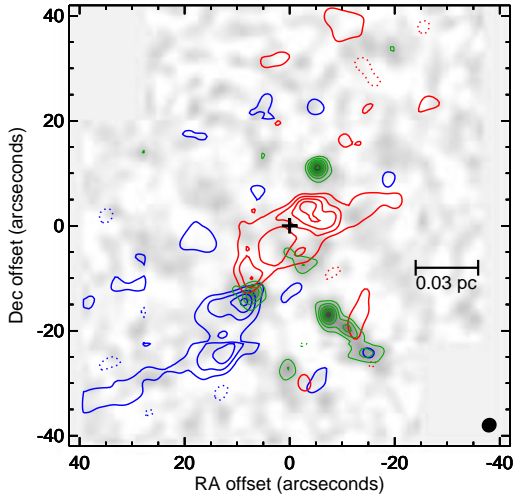}
\caption{\label{fig_12co21_wings}High-velocity redshifted and blueshifted 
integrated \cojtwo\ emission overlaid on the SMA 1.3 mm continuum image 
of the region surrounding HBC722.  The central (0,0) position and the black 
cross mark the position of HBC722 itself.  
The grayscale is inverted and displayed in a linear stretch 
ranging from -1.65 mJy beam$^{-1}$ ($-1\sigma$, white) to 28.05 mJy beam$^{-1}$ 
($17\sigma$, black).  The solid green contours show the continuum intensity 
starting at 3$\sigma$ and increasing by 2$\sigma$, 
where the 1$\sigma$ rms in the continuum image is 1.65 mJy 
beam$^{-1}$.  The blue contours show blueshifted \cojtwo\ emission integrated 
from $-5$ to 1 \kms, while the red contours show redshifted emission integrated 
from 8 to 14 \kms.  The solid blue and red contours start at 2.0 Jy beam$^{-1}$ 
\kms\ and increase by 2.0 Jy beam$^{-1}$ \kms\ whereas the dotted blue and 
red contours start at $-$2.0 Jy beam$^{-1}$ \kms\ and decrease by 2.0 
Jy beam$^{-1}$ \kms.  The synthesized beam size and shape of the \co\ 
observations is shown by the black filled oval in the lower right.}
\end{figure}

The mean number density of each source, $n$, is calculated assuming spherical 
symmetry as

\begin{equation}\label{eq_numberdens}
n = \frac{3}{4 \pi \mu m_H} \frac{M}{r^3_{eff}}
\end{equation}
where $M$ is the mass, $r_{eff}$ is the effective radius\footnote{The effective 
radius is defined as the geometric mean of the semimajor and semiminor axes.}, 
$m_{H}$ is the hydrogen mass, and $\mu$ is the mean molecular weight per free 
particle.  We adopt $\mu = 2.37$ for gas that is 71\% by mass hydrogen, 27\% 
helium, and 2\% metals (Kauffmann et al.~2008).  Again, no uncertainties are 
listed since they are dominated by the uncertainties in mass.

\begin{figure*}
\epsscale{1.0}
\plottwo{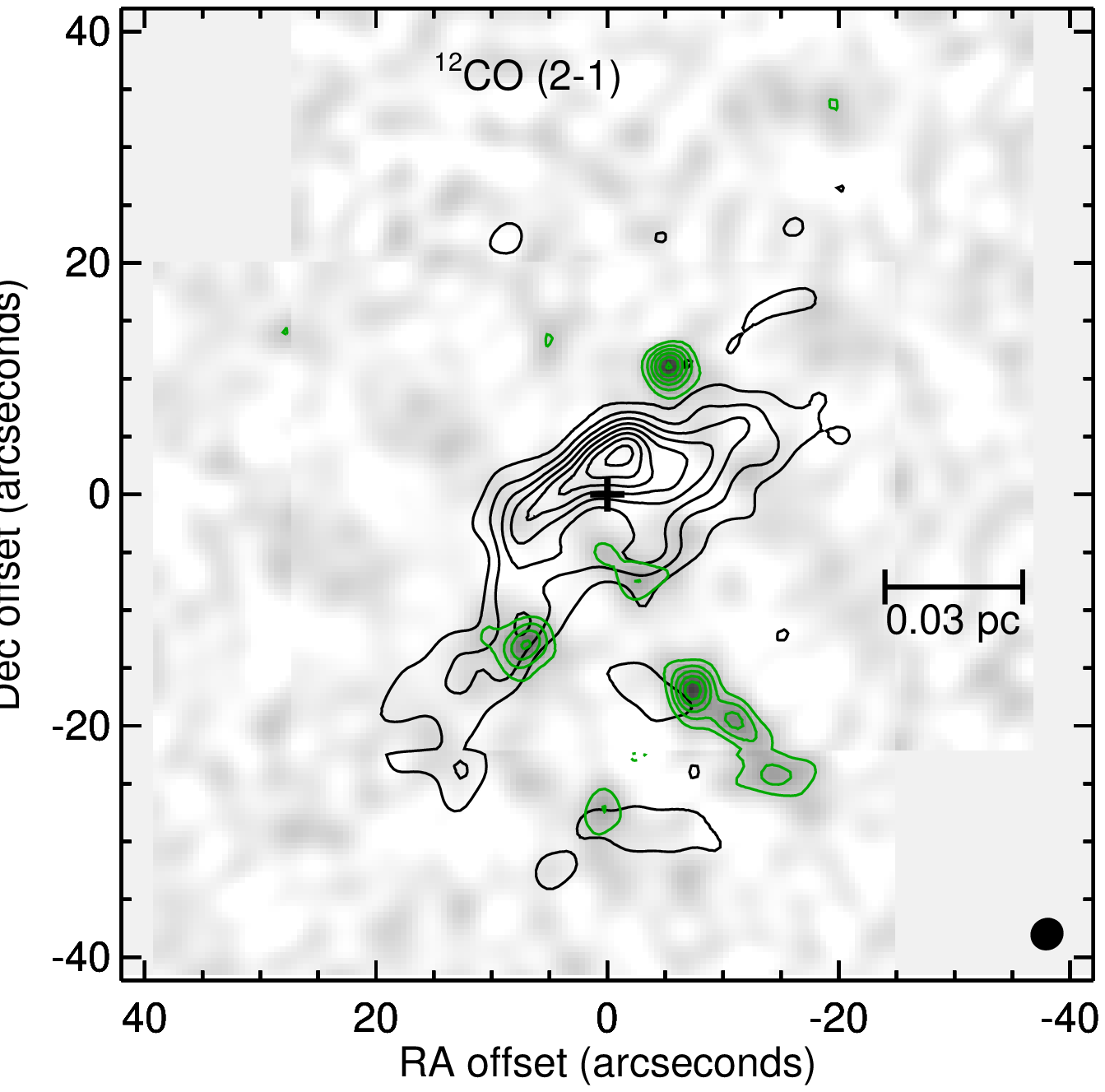}{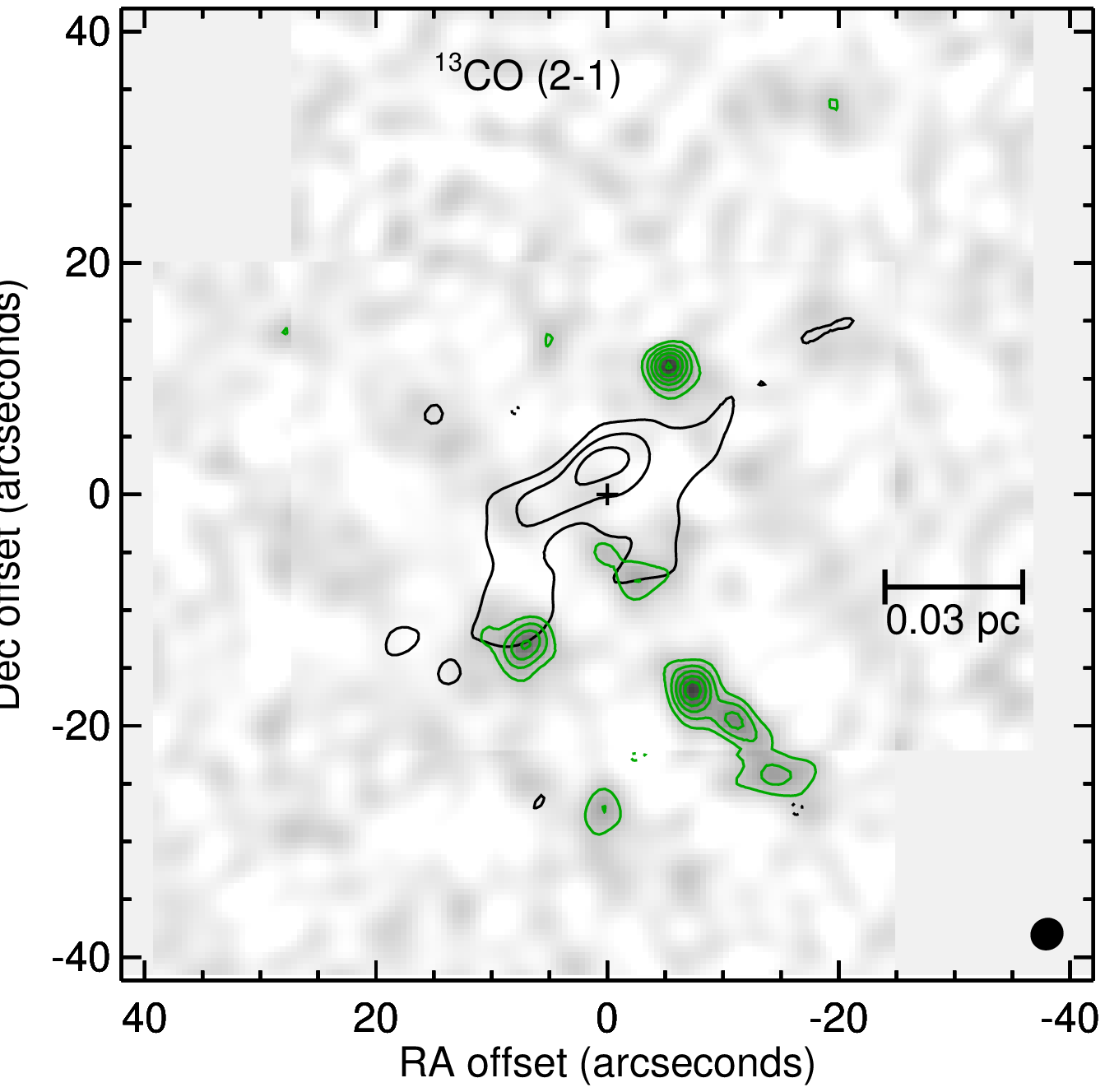}
\caption{\label{fig_12co21_integrated}Low-velocity \co\ (left) and \coo\ 
(right) integrated emission overlaid on the SMA 1.3 mm continuum image of 
the region surrounding HBC722.  The background images and green contours are 
the same as described in Figure \ref{fig_12co21_wings}.  The black contours 
show the \co\ (left) and \coo\ (right) emission integrated from 1 to 8 
\kms.  For \co\ they start at 3 Jy beam$^{-1}$ \kms\ and increase by 3 
Jy beam$^{-1}$ \kms, while for \coo\ they start at 1.2 Jy beam$^{-1}$ \kms\ and 
increase by 1.2 Jy beam$^{-1}$ \kms.  The synthesized beam sizes and shapes of 
the \co\ (left) and \coo\ (right) observations are shown by the black filled 
oval in the lower right.}
\end{figure*}

\begin{figure}
\epsscale{1.0}
\plotone{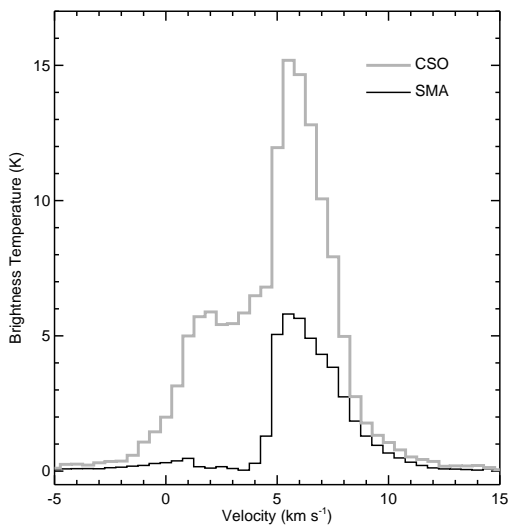}
\caption{\label{fig_12co21_spectra}CSO \cojtwo\ spectrum centered 
on HBC722 from Green et al.~(2011) (gray line) and average SMA 
\cojtwo\ spectrum centered on HBC722 (black line), with the average weighted 
by the CSO beam (assumed to be a Guassian with a 32.5\as\ FWHM).}
\end{figure}

If we treat sources MMS4, MMS5, and MMS6 as a single group of sources and 
average their positions to determine the position of the group, the nearest 
neighbor distance for each of the 5 sources (where MMS4, MMS5, and MMS6 
are now a single source) ranges from $0.03-0.045$ pc.  
This is comparable to the Jeans length of 0.02 -- 0.07 pc for 10 K gas at 
number densities $n = 10^5 - 10^6$ cm$^{-3}$.  However, any further 
interpretation is limited by the fact that we are only able to measure 
projected separations in the plane of the sky.

\subsection{CO}\label{sec_results_co}

Figure \ref{fig_12co21_cmap} displays channel maps of the SMA \cojtwo\ data at 
a velocity resolution of 1 \kms.  Emission is detected between approximately 
$-5$ to 15 \kms.  The lack of emission in the channel centered around 3.25 
\kms\ suggests a cloud systemic velocity around this value since the cloud 
emission is typically fully resolved out in interferometer observations with 
similar \emph{uv} coverage to the data presented here.  
However, with no detections of dense gas tracers, 
including \ntdpjthree\ in these observations, the true systemic velocity 
of HBC722 and other, nearby sources in the region is not well characterized.  
We note that Green et al.~(2011) detected \cojfourteen\ at a velocity of 
6.6 \kms\ in a \emph{Herschel}-HIFI (Heterodyne Instrument for the 
Far-Infrared; de Graauw et al.~2010) observation centered on HBC722, but such 
a high-J transition may be dominated by warm gas in outflows in the region and 
not a good tracer of the true cloud systemic velocity.

In single-dish \cojtwo\ data obtained at the CSO, Green et 
al.~(2011) detected a bipolar molecular outflow centered approximately at 
the position of HBC722, with blueshifted emission extending to the southeast 
and redshifted emission extending to the northwest.  However, with a 32.5\as\ 
beam, they lacked the spatial resolution to definitively identify the driving 
source of this outflow and noted the possibility that none of the detected 
CO emission was related to HBC722 itself.  
Figure \ref{fig_12co21_wings} shows integrated blueshifted and redshifted 
SMA \cojtwo\ emission contours overlaid on the 1.3 mm continuum image, 
integrated over the velocity intervals of $-$5 to 1 \kms\ for the blueshifted 
emission and 8 to 14 \kms\ for the redshifted emission.
The presence of blueshifted emission 
extending to the southeast and redshifted emission extending to the northwest 
is confirmed in the SMA data.  The emission is centered on the continuum source 
MMS3, arguing that it is this source rather than HBC722 that is 
driving the outflow 
detected by Green et al.~and confirmed by these SMA observations.  We also 
note the presence of weak redshifted emission extending to the northwest of 
MMS5 in an L-shaped morphology (see, in particular, the 8.25 \kms\ panel of 
Figure \ref{fig_12co21_cmap}) and weak blueshifted emission extended to the 
southwest of MMS2 (most noticeable in the 0.25 \kms\ panel of Figure 
\ref{fig_12co21_cmap}), possibly indicating the presence of an outflow 
driven by MMS5.

The NW-SE outflow driven by MMS3 shows an S-shaped morphology, similar to 
that observed in several other sources over a range of masses, including 
IRAS 20126+4104 (Shepherd et al.~2000), L1157 (Zhang et al.~2000), and 
RNO43 (Arce \& Sargent 2005).  This type of morphology is typically explained 
by an outflow axis that precesses with time.  
Such precession is generally thought to arise 
from either tidal interactions between the disk of the source driving the 
outflow and companion sources or from anisotropic accretion events 
(e.g., Shepherd et al.~2000).  A similar morphology, in particular an apparent 
change in the axis of the red lobe from one oriented more north-south to 
one oriented more east-west as distance from the driving source increases, is 
also hinted at in the single-dish data presented by Green et al.~(2011).  
However, as discussed in more detail below, 
these SMA observations are only recovering the densest components of a very 
extended emission morphology, leading to significant uncertainty in the true 
underlying morphology of this outflow.  We conclude that there is tentative 
but unconfirmed evidence that the axis of the outflow driven by MMS3 is 
precessing.

High-velocity outflowing gas from MMS3 is not the only source of CO emission in 
the SMA map.  Inspection of Figure \ref{fig_12co21_cmap} clearly shows 
substantial \cojtwo\ emission between 1 and 8 \kms, the lower bounds for the 
blueshifted and redshifted emission, respectively.  Figure 
\ref{fig_12co21_integrated} plots both \cojtwo\ and \coojtwo\ emission 
contours integrated between $1-8$ \kms.  Most of the emission is located along 
the NW-SE axis of the outflow driven by MMS3 and likely arises from a 
lower-velocity component of this outflow.  However, MMS1, MMS2, 
and HBC722 are all located near the strongest emission, and neither the 
spatial resolution nor the emission morphology clearly indicate whether all of 
this emission is due to the outflow driven by MMS3 or if additional weak, 
low-velocity outflows are present in this region.

Analysis of the outflow(s) in this region is further complicated by 
the fact that, with minimum baselines of $\sim 5$ k$\lambda$ (corresponding 
to spatial scales of 25700 AU at a distance of 520 pc) and less than 
2\% of the total \emph{uv} pointings located at projected baselines 
$< 10$ k$\lambda$ (corresponding to spatial scales of 12900 AU at a distance of 
520 pc), these observations are not sensitive to extended outflow emission.  
The \cojtwo\ emission shown in Figures \ref{fig_12co21_cmap} -- 
\ref{fig_12co21_integrated} likely only represents the densest, most compact 
components of a more extended emission morphology.  This statement is 
confirmed by Figure 2 of Green et al.~(2011), which clearly shows large-scale 
\cojtwo\ emission extending over $>$100\as\ (52000 AU at the distance of 520 
pc).  It is further confirmed by Figure \ref{fig_12co21_spectra}, which 
compares the CSO \cojtwo\ spectrum at the position of HBC722 from Green 
et al.~(2011) to the average SMA \cojtwo\ spectrum centered on HBC722, where 
the average is weighted by the CSO beam (assumed to be a Gaussian with a 
32.5\as\ FWHM) and is calculated in practice in the image plane by adding 
together all emission within 32.5\as\ of HBC722 and down-weighting based on 
the distance from HBC722 with the assumed CSO beam.  
Comparing the CSO and SMA spectra show that at least 50\% 
of the emission is resolved out over essentially all velocities for which 
CO emission is detected, and more than 90\% is resolved out between $1-4$ 
\kms.

\begin{figure*}
\epsscale{1.0}
\plotone{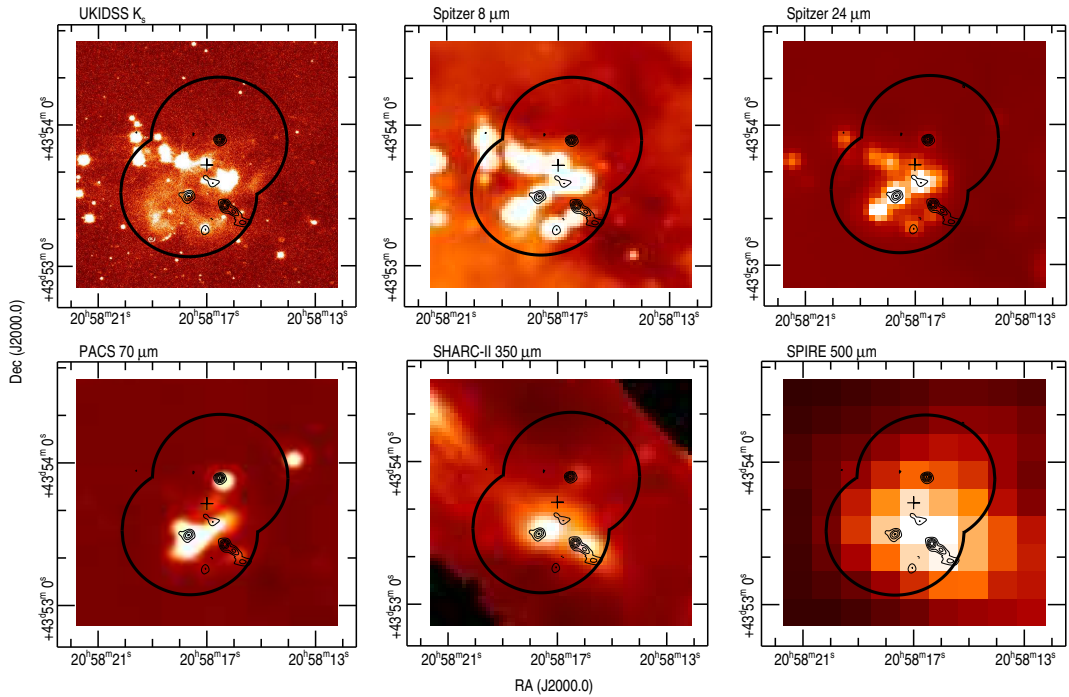}
\caption{\label{fig_cont_overlay}SMA 1.3mm continuum intensity contours 
overlaid on infrared and submillimeter images of the region surrounding 
HBC722.  The panels show, from left-to-right and top-to-bottom, UKIDSS 
K$_{\rm s}$ band, \emph{Spitzer} 8 \um, \emph{Spitzer} 24 \um, \emph{Herschel} 
70 \um, SHARC-II 350 \um, and \emph{Herschel} 500 \um\ images.  The cross in 
each panel marks the position of HBC722 itself, and the thick black lines show 
the primary beam of the SMA centered on the two pointings in the mosaic.  The 
thin black lines show the SMA 1.3 mm continuum intensity contours, with the 
solid lines starting at 3$\sigma$ and increasing by 2$\sigma$ and the dashed 
lines starting at $-3\sigma$ and decreasing by $2\sigma$, where the 1$\sigma$ 
rms in the continuum image is 1.65 mJy beam$^{-1}$.  Pointing offsets on the 
order of a few arcseconds are visible between the 
SMA continuum emission and \emph{Herschel} and SHARC-II continuum emission; 
these offsets are within the pointing uncertainties of \emph{Herschel} and 
SHARC-II and are much smaller than the single-dish beams, thus we consider 
them to be negligible.}
\end{figure*}

Ultimately, we confirm the presence of at least one NW-SE outflow in the 
vicinity of HBC722, as suggested by Green et al.~(2011).  The morphology of 
this outflow suggests MMS3 rather than HBC722 as the most likely driving 
source, and suggests that this outflow may be precessing.  
There is also an extremely tentative detection of a weak outflow 
driven by MMS5.  We do not find any conclusive evidence that HBC722 
itself is driving an outflow.  Given the uncertainties in the systemic 
velocity of the region and in the true morphology of the CO emission due to 
resolving out extended emission, we are unable to completely rule out the 
possibility of additional outflows in the region, particularly at low 
velocities relative to the uncertain core systemic velocity.  
Indeed, we note that there is \co\ and \coo\ emission spatially coincident 
with HBC722, but lack sufficient information to determine whether its origin 
lies in one or more outflows or simply in ambient cloud emission.  Future 
observations that provide both higher spatial resolution \emph{and} better 
sensitivity to extended emission are required to fully analyze the outflows in 
the vicinity of HBC722.

\section{Discussion}\label{sec_discussion}

\subsection{Continuum Sources}\label{sec_discussion_sources}

Figure \ref{fig_cont_overlay} displays infrared and submillimeter images of 
the HBC722 environment, including an UKIDSS (UKIRT [United Kingdom Infrared 
Telescope] Infrared Deep Sky Survey) $K_{\rm S}$ image, a \emph{Spitzer} 8 \um\ 
image from Guieu et al.~(2009), a \emph{Spitzer} 24 \um\ image from Rebull et 
al.~(2011), and a \emph{Herschel} 70 \um\ image, SHARC-II 350 \um\ image, and 
\emph{Herschel} 500 \um\ image 
from Green et al.~(2011).  Overplotted in black are the SMA 1.3 mm continuum 
intensity contours and the primary beams of the SMA pointings.  A version of 
this Figure without the SMA 1.3 mm continuum intensity contours was 
previously presented by Green et al.~(2011).  

\begin{figure*}
\epsscale{1.0}
\plotone{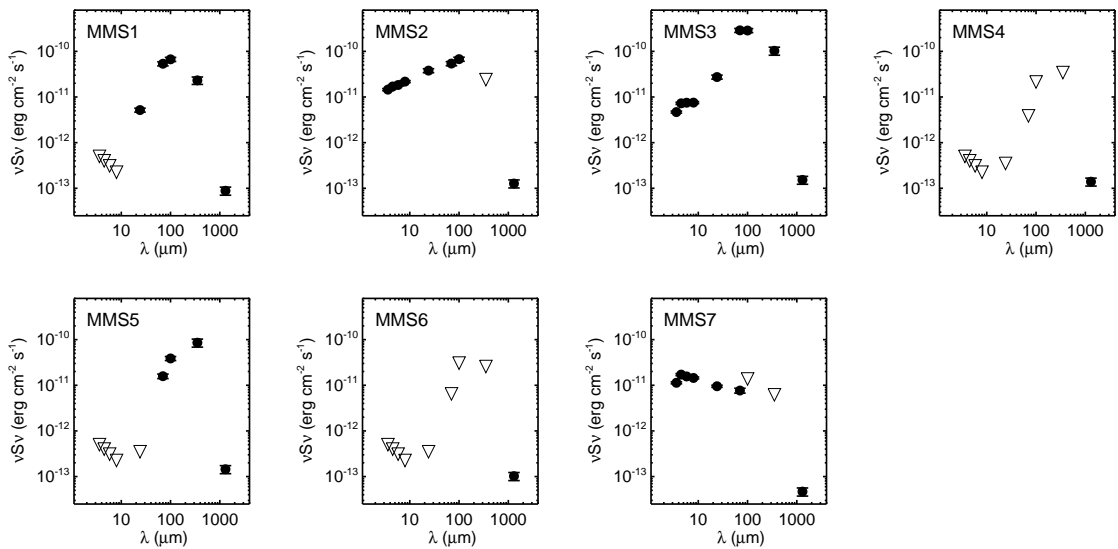}
\caption{\label{fig_seds}Spectral Energy Distributions (SEDs) for each of 
the seven detected continuum sources, consisting of \emph{Spitzer} $3.6-24$ 
\um\ data, \emph{Herschel} 70 and 100 \um\ data, SHARC-II 350 \um\ data, 
and the SMA 1.3 mm continuum data (see text for details).  Detections are 
plotted as filled circles with error bars and upper limits are plotted as open 
triangles.  Each panel is labeled with the corresponding source.}
\end{figure*}

Inspection of Figure \ref{fig_cont_overlay} shows that most of the seven 
detected SMA continuum sources are associated with sources at other 
wavelengths.  A detailed discussion of each individual continuum source and 
its associations with sources at other wavelengths is given below.  Table 
\ref{tab_photometry} lists, for each source, \emph{Spitzer} photometry at 
$3.6-24$ \um\ from Guieu et al.~(2009) and Rebull et al.~(2011), 
\emph{Herschel} 70 and 100 \um\ photometry from Green et al.~(2011), 
SHARC-II 350 \um\ photometry from Green et al.~(2011), and SMA 1.3 mm 
continuum photometry from this work.  Figure \ref{fig_seds} plots SEDs of all 
seven continuum sources, including both detections and upper limits (see 
below for more details). 

\begin{deluxetable*}{lccccccc}
\tabletypesize{\scriptsize}
\tablewidth{0pt}
\tablecaption{\label{tab_photometry}Infrared and (Sub)millimeter Photometry of Continuum Sources}  
\tablehead{
\colhead{Wavelength} & \colhead{MMS1} & \colhead{MMS2} & \colhead{MMS3} & \colhead{MMS4} & \colhead{MMS5} & \colhead{MMS6} & \colhead{MMS7} \\
\colhead{(\um)} & \colhead{S$_{\nu}$ (mJy)} & \colhead{S$_{\nu}$ (mJy)} & \colhead{S$_{\nu}$ (mJy)} & \colhead{S$_{\nu}$ (mJy)} & \colhead{S$_{\nu}$ (mJy)} & \colhead{S$_{\nu}$ (mJy)} & \colhead{S$_{\nu}$ (mJy)}
}
\startdata
3.6                   & $<$0.6         & 17.4 $\pm$ 0.9 & 5.61 $\pm$ 0.28   & $<$0.6        & $<$0.6           & $<$0.6         & 13.6 $\pm$ 0.68 \\
4.5                   & $<$0.6         & 25.5 $\pm$ 1.3 & 10.9 $\pm$ 0.55   & $<$0.6        & $<$0.6           & $<$0.6         & 25.7 $\pm$ 1.3  \\
5.8                   & $<$0.6         & 35.4 $\pm$ 1.8 & 14.5 $\pm$ 0.72   & $<$0.6        & $<$0.6           & $<$0.6         & 30.2 $\pm$ 1.5  \\
8.0                   & $<$0.6         & 58.0 $\pm$ 2.9 & 20.1 $\pm$ 1.0    & $<$0.6        & $<$0.6           & $<$0.6         & 38.6 $\pm$ 1.9  \\
24                    & 41.1 $\pm$ 4.1 & 303 $\pm$ 30   & 219 $\pm$ 22      & $<$2.8        & $<$2.8           & $<$2.8         & 76.2 $\pm$ 3.8  \\
70\tablenotemark{a}   & 1250 $\pm$ 130 & 1260 $\pm$ 130 & 6730 $\pm$ 680    & $<$90         & 369 $\pm$ 42     & $<$151         & 179 $\pm$ 22    \\
100\tablenotemark{a}  & 2230 $\pm$ 230 & 2250 $\pm$ 230 & 9530 $\pm$ 960    & $<$712        & 1290 $\pm$ 130   & $<$1020        & $<$458          \\
350\tablenotemark{b}  & 2700 $\pm$ 500 & $<$2800        & 12000 $\pm$ 2400  & $<$4000       & 10000 $\pm$ 2000 & $<$3000        & $<$720          \\
1300                  & 38.3 $\pm$ 7.7 & 55.1 $\pm$ 11  & 66.1 $\pm$ 13     & 60.5 $\pm$ 12 & 62.7 $\pm$ 13   & 44.3 $\pm$ 8.9 & 20.3 $\pm$ 4.1 
\enddata
\tablenotetext{a}{All flux densities and upper limits are derived from the \emph{Herschel} 70 and 100 \um\ maps from Green et al.~(2011).  The flux densities are calculated in 10\as\ diameter apertures, and the upper limits are described in the text.}
\tablenotetext{b}{All SHARC-II 350 \um\ flux densities are measured in 20\as\ diameter apertures (see Wu et al.~2007 for more details on SHARC-II aperture photometry and calibration).}
\end{deluxetable*}

\begin{deluxetable}{lccc}
\tabletypesize{\scriptsize}
\tablewidth{0pt}
\tablecaption{\label{tab_evolindic}Evolutionary Indicators}  
\tablehead{
\colhead{}       & \colhead{}         & \colhead{\tbol} & \colhead{\lbol}   \\
\colhead{Source} & \colhead{$\alpha$} & \colhead{(K)}   & \colhead{(\lsun)}}
\startdata
MMS1 & \nodata & 39      & 1.1     \\
MMS2 & 0.49    & 147     & 1.5     \\
MMS3 & 0.76    & 52      & 5.9     \\
MMS4 & \nodata & \nodata & \nodata \\
MMS5 & \nodata & 15      & 1.4     \\
MMS6 & \nodata & \nodata & \nodata \\
MMS7 & $-$0.20 & 328     & 0.3     
\enddata
\end{deluxetable}

Table \ref{tab_evolindic} presents three evolutionary indicators 
for each source calculated from the SEDs tabulated in Table 
\ref{tab_photometry} and plotted in Figure \ref{fig_seds}:  the infrared 
spectral index ($\alpha$), the bolometric temperature (\tbol), and the 
bolometric luminosity (\lbol).  As first defined by Lada \& Wilking (1984) 
and Lada (1987), $\alpha$ is the infrared slope in log space of $\nu S_{\nu}$ 
vs.~$\nu$ and is used to classify sources into different evolutionary 
stages (see Evans et al.~[2009] for a recent review of classification via 
$\alpha$).  In this study we calculate $\alpha$ with a linear least-squares 
fit to all available \emph{Spitzer} photometry between $3.6-24$ \um.  \tbol\ is 
defined as the temperature of a blackbody with the same flux-weighted mean 
frequency as the source SED (Myers \& Ladd 1993) and provides an alternative 
classification method to $\alpha$ (Chen et al.~1995; Evans et al.~2009).  
We calculate both \tbol\ and \lbol\ by using the trapezoid 
rule to integrate over the finely sampled SEDs; a detailed description of 
the implementation of this method and resulting errors due to the finite 
sampling of the observed SEDs is given in Appendix B of Dunham et al.~(2008).  
An additional error is introduced by the fact that the SMA 1.3 mm continuum 
data resolve out some of the emission from the extended core and are thus 
lower limits to the true flux densities at this wavelength, artifically 
steepening the far-infrared and submillimeter slope of the SEDs.  The 
magnitude of the error introduced depends on both the amount of emission 
recovered by the SMA and the spectral shape of each source; a very 
conservative estimate that only 1\% of the emission is recovered leads to 
underestimates in \lbol\ by less than a factor of 2 for all sources, and 
less than 50\% for all but MMS7.  Thus, while we caution that the calculated 
\lbol\ may underestimate the true values, the magnitude of these underestimates 
are likely comparable to or less than the other errors discussed by Dunham et 
al.~(2008).

Finally, we note that there are three other infrared sources identified as 
\emph{Spitzer}-detected Young Stellar Objects by Guieu et al.~(2009) and 
Rebull et al.~(2011) covered by our SMA observations, and several other 
infrared sources covered that are not identified as Young Stellar Objects and 
are thus likely background or foreground stars.  As our focus is on providing 
a high spatial resolution view of the millimeter emission in the vicinity of 
HBC722 we do not discuss these sources further except to note that their 
non-detections in our SMA 1.3 mm continuum observations suggest 
similar upper limits to the masses of any circumstellar disks surrounding 
these objects as for HBC722 itself, which we discuss below in \S 
\ref{sec_discussion_hbc722}.

\subsubsection{MMS1}

MMS1 is associated with the \emph{Spitzer} infrared source SST 
J205816.56$+$435352.9 from Guieu et al.~(2009) and Rebull et al.~(2011).  It 
is detected by \emph{Spitzer} only at 24 \um\ with the flux density listed in 
Table 
\ref{tab_photometry}.  The \emph{Spitzer} $3.6-8$ \um\ upper limits listed 
are taken from Guieu et al.~(2009); they note that their 90\% completeness 
limits increase from about 0.3 mJy at 3.6 \um\ to 0.6 mJy at 8 \um\ but do not 
specifically list these limits for 4.5 and 5.8 \um, thus we conservatively 
take the upper limits to be 0.6 mJy in all four bands.  MMS1 is also 
associated with a \emph{Herschel} source detected at 70 and 100 \um\ from 
Green et al.~(2009); all other \emph{Herschel} wavelengths are of too low 
spatial resolution ($18-35$\as\ at $160-500$ \um) to accurately separate 
the 7 continuum sources and are not considered in this study.  The 
70 and 100 \um\ photometry presented in Table \ref{tab_photometry} 
is calculated with 10\as\ diameter apertures chosen as the best compromise 
between including as much source flux as possible and excluding flux from 
other, nearby sources.  No aperture or color corrections are applied since 
neither the true spatial profile 
of the emission nor the underlying spectral shape of any 
source is known.  Finally, MMS1 is also associated with a SHARC-II 350 \um\ 
source.  The photometry presented in Table \ref{tab_photometry} is calculated 
in a 20\as\ diameter aperture following the method described by Wu et 
al.~(2007).

The SED of MMS1 resembles that of a deeply embedded protostar.  We are unable 
to calculate $\alpha$ with only one \emph{Spitzer} detection, but the upper 
limits are consistent with a positive $\alpha$ indicative of a Class 0/I 
object.  We calculate \tbol\ $=39$ K, placing MMS1 in Class 0, consistent 
with the above statements.  We thus conclude that MMS1 is a deeply embedded 
Class 0 protostar.  As described in \S \ref{sec_results_co} above, 
we do not detect clear signatures of an outflow driven by MMS1 but do note that 
there may be additional outflows in this region that are not fully separated 
spatially and kinematically by these data.

\subsubsection{MMS2}

MMS2 is associated with the \emph{Spitzer} infrared source SST 
J205816.8$+$435335.6 from Guieu et al.~(2009) and Rebull et al.~(2011) and 
is detected at all five \emph{Spitzer} wavelengths from $3.6-24$ \um.  
It is also detected at 70 and 100 \um\ with \emph{Herschel}.  The photometry 
is again calculated 
in 10\as\ diameter apertures and is uncertain since the apertures do not 
fully capture the emission that extends to the northwest but also partially 
overlap with the emission from the brighter MMS3.  More accurate photometry 
at these wavelengths will require higher-resolution observations.  There is 
no clear SHARC-II 350 \um\ source, but the position of MMS2 overlaps with 
bright emission from MMS3 and MMS5.  We calculate the upper limit as the 
flux in one beam at the position of MMS2 from the other, nearby sources.

The SED of MMS2 resembles that of a Class I protostar more evolved and less 
deeply embedded than MMS1.  The calculated values of $\alpha$ and \tbol\ 
(0.49 and 147 K, respectively) both classify MMS2 as Class I, confirming this 
statement.

\subsubsection{MMS3}

MMS3 is associated with the \emph{Spitzer} infrared source SST 
J205817.7$+$435331.1 from Guieu et al.~(2009) and Rebull et al.~(2011) and 
is detected at $3.6-24$ \um\ with \emph{Spitzer}.  It is also detected 
at 70 and 100 \um\ with \emph{Herschel} and 350 \um\ with SHARC-II and is 
the brightest source in the region at these wavelengths.  The photometry at 
70, 100, and 350 \um\ is calculated with apertures and methods identical to 
MMS1 above.  The true flux densities at these wavelengths are likely higher 
than the values listed in Table \ref{tab_photometry} since the apertures do 
not include all of the extended emission, but larger apertures are not 
feasible since they would overlap with other, nearby sources.

As with MMS1, the SED of MMS3 resembles that of a deeply embedded protostar.  
The calculated values of both $\alpha$ (0.76; in the Class 0/I category) and 
\tbol\ (52 K; in the Class 0 category) are consistent with this observation.  
As discussed in \S \ref{sec_results_co} above, MMS3 is driving a NW-SE outflow 
detected both by the SMA \cojtwo\ observations presented here and single-dish 
\cojtwo\ observations presented by Green et al.~(2011).  The axis of this 
outflow may be precessing over time.

\subsubsection{MMS4}

MMS4 is not associated with a \emph{Spitzer} infrared source at $3.6-24$ 
\um, it is not associated with a \emph{Herschel} infrared source at 70 and 
100 \um, and it is not associated with a SHARC-II 350 \um\ submillimeter 
source.  The upper limits for $3.6-8$ \um\ are again taken from Guieu et 
al.~(2009) as described above for MMS1.  For 24 \um, we take the upper 
limit to be the point at which the source count histogram presented in 
Figure 9 of Rebull et al.~(2011) turns over, which we estimate to be at a 
magnitude of 8.5 (corresponding to a flux density of 2.8 mJy).  While there 
is no \emph{Herschel} 70 or 100 \um\ source or SHARC-II 350 \um\ source, there 
is emission at the position of MMS4 from the brighter nearby sources (MMS3 
and MMS5).  Thus, similar to MMS2 above, we calculate the upper limits as 
the flux in one beam at the position of MMS4 from other, nearby sources.

With no detections of a compact, infrared source in any of the \emph{Spitzer} 
bands, and also no detections at $70-350$ \um, MMS4 is possibly a starless 
core heated only externally and thus too faint to detect at 350 \um\ above the 
emission from nearby, brighter sources.  The 350 \um\ upper limit of 4 Jy is 
consistent with this statement since starless cores and cores containing very 
low luminosity protostars are typically less than 1 -- 2 Jy at this 
wavelength (Wu et al.~2007).  However, recent work suggests that 
detections of starless cores with current interferometers are extremely rare 
since starless cores are not yet very centrally condensed and are thus fully 
resolved out (Schnee et al.~2010; Offner et al.~2012).  
If MMS4 is indeed a starless core, the SMA detection 
indicates it may be very evolved and close to the onset of star formation.  
Alternatively, star formation may have already begun, with the core harboring 
a very young, very low luminosity protostar or first hydrostatic core.  
Confirmation would require either the detection of a very faint infrared 
source below the detection limits of the data considered here, such as the 
70 \um\ detection of the source Per-Bolo 58 presented by Enoch et al.~(2010), 
or the detection of a molecular outflow driven by this core, such as the 
detections of outflows from cores previously believed to be starless 
presented by Chen et al.~(2010), Dunham et al.~(2011), Pineda et al.~(2011), 
and Schnee et al.~(2012).  No such outflow is detected in our \cojtwo\ 
observations, but this topic should be revisited with future observations 
providing higher sensitivity and higher spatial resolution.

\subsubsection{MMS5}

MMS5 is not associated with a \emph{Spitzer} infrared source at $3.6-24$ 
\um, and the upper limits listed in Table \ref{tab_photometry} are determined 
as described above.  MMS5 is associated with a source detected at 70 and 
100 \um\ with \emph{Herschel} and 350 \um\ with SHARC-II, and the photometry 
at these wavelengths is calculated with apertures and methods identical to 
MMS1 above.  The SED of MMS5 resembles that of a Class 0 protostar too deeply 
embedded to be detected in the mid-infrared with \emph{Spitzer}.  With no such 
detections we are unable to calculate $\alpha$, but calculate a value for 
\tbol\ (15 K) consistent with that of a Class 0 source.  As noted in \S 
\ref{sec_results_co}, there is weak redshifted emission extending 
to the northwest of MMS5 and weak blueshifted emission to the southeast.  
These weak features may be due to an 
outflow driven by MMS5, but higher sensitivity \co\ observations are required 
to confirm this tentative outflow.

\subsubsection{MMS6}

Similar to MMS4, MMS6 is not associated with a \emph{Spitzer} infrared source 
at $3.6-24$ \um, it is not associated with a \emph{Herschel} infrared source 
at 70 and 100 \um, and it is not associated with a SHARC-II 350 \um\ 
submillimeter source.  The $3.6-24$ \um\ upper limits listed in Table 
\ref{tab_photometry} are determined as described above.  The 70, 100, and 350 
\um\ upper limits are calculated as the flux in one beam at the position of 
MMS6 from other, nearby sources.  The same discussion presented above for the 
evolutionary status of MMS4 also applies for MMS6; it is likely either an 
evolved starless core close to the onset of star formation or a very young, 
very low luminosity protostar or first hydrostatic core.  As with MMS4, we do 
not detect any evidence for an outflow driven by MMS6.

\subsubsection{MMS7}

MMS7 is associated with the \emph{Spitzer} infrared source SST 
J205817.06$+$435316.1 from Guieu et al.~(2009) and Rebull et al.~(2011) and 
is detected at all five \emph{Spitzer} wavelengths.  It is also detected at 
70 \um\ with \emph{Herschel}, and the photometry presented in Table 
\ref{tab_photometry} is calculated with an aperture and methods identical to 
MMS1 above.  It is not detected at either 100 \um\ with \emph{Herschel} or 
350 \um\ with SHARC-II; upper limits are again calculated as the flux in one 
beam at the position of MMS7 from other, nearby sources.  The SED of MMS7 
resembles that of a young stellar object (YSO) surrounded by a circumstellar 
disk but no longer embedded within a dense core; the calculated values 
of $\alpha$ ($-$0.20; in the Flat Spectrum category) and \tbol\ (328 K; near 
the Class I/II boundary) confirm this assessment.

\subsection{Evolutionary Status of HBC722}\label{sec_discussion_hbc722}

Prior to outburst HBC722 was regarded as a Class II T Tauri star with a 
spectral type of K7$-$M0, a mass of $\sim 0.5-0.6$ \msun, a visual extinction 
of 3.4 magnitudes, an infrared spectral index of $-0.77$, and a bolometric 
luminosity of 0.85 \lsun\ (Cohen \& Kuhi 1979; \kospal\ et al.~2011; Miller 
et al.~2011).  As noted in \S \ref{sec_results_continuum}, HBC722 itself is 
not detected in our SMA 1.3 mm continuum observations down to a $3\sigma$ 
upper limit of 5 mJy beam$^{-1}$.  Under the same assumptions as discussed 
above in \S \ref{sec_results_continuum}, this corresponds to an upper limit of 
0.02 \msun\ for the disk mass (lower if the disk is warmer than 30 K) 
and an upper limit of 3\% -- 4\% for the ratio of disk to stellar mass (again, 
lower if the disk is actually warmer than 30 K).  
In a large submillimeter survey of circumstellar 
disks around young stars, Andrews \& Williams (2005) report a mean disk mass 
of 0.005 \msun\ with a large dispersion of $\sim$0.5 dex, and a median ratio 
of disk to stellar mass of 0.5\%.  Thus, our observations do not rule out the 
presence of a typical mass disk.

In a typical FUor flare with an accretion rate of 10$^{-4}$ \msun\ yr$^{-1}$ and 
a duration of 100 yr, up to 0.01 \msun\ of mass can accrete from the disk 
onto the protostar.  For HBC722, however, \kospal\ et al.~(2011) calculated a 
burst accretion rate of only 10$^{-6}$ based on their measured \lbol\ during 
the burst\footnote{This measurement is rather uncertain due to the variability 
of the outburst brightness since the initial flare in 2010 (see \S 
\ref{sec_hbc722}) and the fact that \kospal\ et al.~lacked photometry during 
the burst at $\lambda > 10$ \um, although the latter point is mitigated 
by the fact that the \emph{Herschel} 70 \um\ flux 
density of 412 mJy measured during the burst and reported by Green et 
al.~(2011) is generally consistent with the assumptions made by \kospal\ 
et al.~to extrapolate beyond 10 \um.}.  Our derived upper limit for the disk 
mass of 0.02 \msun\ is thus consistent with providing a sufficient mass 
reservoir to support 
the observed outburst in HBC722 unless the accretion rate is several orders 
of magnitude higher than estimated by \kospal\ et al.~and/or the burst duration 
is much longer than the typical 100 yr.  Even if one of these cases were true, 
there is possibly gas remaining in the vicinity of HBC722 that could still 
accrete onto the star+disk system and power the burst, 
as suggested by \coojtwo\ and 
\emph{Herschel} far-infrared continuum emission spatially coincident with 
HBC722.  Further constraints on the amount of circumstellar mass available 
to accrete onto HBC722 and the likelihood of sufficient mass remaining to 
power further bursts beyond 
the current one require deeper millimeter continuum data probing to lower 
disk masses and higher-resolution far-infrared and submillimeter continuum 
data observed during the burst to better determine the burst luminosity 
and implied accretion rate.  For the former, the high sensitivity 
of full-science ALMA operations will be the ideal facility despite the high 
declination of HBC722 ($+$44\degree) since, according to the ALMA sensitivity 
calculator\footnote{Available at 
https://almascience.nrao.edu/call-for-proposals/sensitivity-calculator}, even 
a short, 30 minute track with 50 antennas will improve the mass sensitivity 
by a factor of 100.  For the latter, the upcoming 25-m submillimeter telescope 
CCAT will be of particular value assuming the burst is still in progress when 
CCAT begins science operations (currently expected in 2015 -- 2017; Radford et 
al.~2009; Sebring 2010).

\section{Summary}\label{sec_summary}

In this paper we have presented 230 GHz Submillimeter Array continuum and 
molecular line observations of 
the newly discovered FUor candidate HBC722.  We summarize our main results 
as follows.

\begin{enumerate}
\item Seven 1.3 mm continuum sources are detected in the vicinity of HBC722; 
none are HBC722 itself.  We compile infrared and submillimeter continuum 
photometry of each source from previous studies and conclude that three are 
Class 0 embedded protostars, one is a Class I embedded protostar, one is a 
Class I/II transition object, and two are either starless cores or very 
young, very low luminosity protostars or first hydrostatic cores.
\item A northwest-southeast outflow is detected in the \cojtwo\ 
observations.  This outflow is centered on and thus likely driven by MMS3, 
one of the Class 0 sources detected in the 1.3 mm continuum data, and its axis 
may be precessing.  This outflow 
detection is consistent with a similar outflow detected in low-resolution, 
single-dish \cojtwo\ observations presented by Green et al.~(2011).  
Our higher spatial resolution confirms that HBC722 is not the driving source.
\item There is no conclusive evidence that HBC722 itself is driving an 
outflow, although we caution that higher spatial resolution, better sensitivity 
to extended emission, and better determinations of the systemic velocities of 
the sources in the vicinity of HBC722 are needed to fully evaluate the 
kinematics of the \cojtwo\ gas in this region.
\item The non-detection of HBC722 in the 1.3 mm continuum observations suggests 
an upper limit of 0.02 \msun\ for the mass of the circumstellar disk, 
consistent with typical T Tauri disks.  This upper limit is consistent with 
a disk that provides sufficient 
mass to power the burst.  Future observations are needed to further study the 
actual amount of circumstellar mass available to accrete onto HBC722 and 
the likelihood of sufficient mass remaining to power additional bursts beyond 
the current one.
\end{enumerate}

We have noted in the text several future observations that are needed in order 
to better disentangle the millimeter emission in this complicated environment 
and better determine the properties and evolutionary status of HBC722.

\acknowledgements
The authors express their gratitude to L.~Rebull and S.~Guieu for providing 
their \emph{Spitzer} images of HBC722.  We thank Neal Evans for reading 
a draft in advance of publication and providing helpful comments.  
This work is based primarily on observations obtained with The Submillimeter 
Array, a joint project between the Smithsonian Astrophysical Observatory and 
the Academia Sinica Institute of Astronomy and Astrophysics and funded by the 
Smithsonian Institution and the Academia Sinica.  This research has made use 
of NASA's Astrophysics Data System (ADS) Abstract Service, the IDL Astronomy 
Library hosted by the NASA Goddard Space Flight Center, and the SIMBAD 
database operated at CDS, Strasbourg, France.  Support for this work was 
provided by the NSF through grant AST--0845619 to HGA.  TvK is grateful for
 the Joint ALMA Observatory for providing the facilities for his scientific 
research.

\end{document}